\documentclass[12pt,a4paper,leqno]{article}
\usepackage{amsmath}
\numberwithin{equation}{section}
\title{On traveling waves in  lattices: The case of Riccati lattices}
\author{Zlatinka I. Dimitrova}
\date{"G. Nadjakov" Institute of Solid State Physics, Bulgarian Academy of Sciences, Blvd.
Tzarigradsko Chausse 72, 1784, Sofia, Bulgaria \\
e-mail: zdim@phys.bas.bg \vskip1cm
[Received 17 April 2012; Accepted 14 May 2012]}
\begin{document}
\maketitle
\begin{abstract}
The method of simplest equation is applied for analysis of a class of lattices described by 
differential-difference equations that admit traveling-wave solutions constructed on the
basis of the solution of the Riccati equation. We denote such lattices as Riccati lattices. 
We search for Riccati lattices within two classes of lattices: generalized Lotka - Volterra 
lattices and generalized Holling lattices. We show that from the class of generalized Lotka - 
Volterra lattices only the Wadati lattice belongs to the class of Riccati lattices. Opposite 
to this many lattices from the Holling class are Riccati lattices. We construct exact traveling 
wave solutions on the basis of the solution of Riccati equation for three  members of the
class of generalized Holing lattices. 
\end{abstract}
\begin{flushleft}
{\bf Key words:} 
\end{flushleft}
nonlinear differential-difference equations,  method of simplest equation, exact traveling-wave solutions,
Lotka - Volterra lattices, Holling lattices, Wadati lattice, Riccati lattices
\section{Introduction}
Nonlinear models are used extensively in the research on complex systems 
\cite{murr} - \cite{frank}. In many cases, the models consist of nonlinear partial 
differential equations and  it is of great interest to obtain exact analytical solutions of 
these nonlinear PDEs. Such solutions are useful as 
initial conditions in the  process of obtaining of numerical solutions. In addition, the exact 
solutions describe important classes of waves and processes in the investigated systems. The 
researches based on  nonlinear PDEs increase steadily  and now they are much applied in the 
theory of solitons \cite{ac}-\cite{kivshar},  biology \cite{scott02}, theory of dynamical 
systems, chaos  theory and ecology \cite{perko} - \cite{vit09a},  hydrodynamics and theory of 
turbulence \cite{temam} - \cite{ott4} , in the mathematical social dynamics \cite{vdm, vp1}, 
etc.  The inverse scattering transform and the method of
Hirota \cite{gardner} - \cite{hirota} are famous methods for obtaining exact soliton 
solutions of various NPDEs. In addition, in the last several years approaches for obtaining 
exact special solutions of nonlinear PDE have been developed, too  \cite{kudr90} - \cite{vit11}.
Numerous exact solutions of many equations have been obtained by means of these approaches
 such as for an example the Kuramoto-Shivasinsky equation 
\cite{vdk10}, \cite {k08a} -\cite {kudr05x}, sine - Gordon equation \cite{lou} - \cite
{nakamura}, equations, connected to the models of population dynamics \cite{zeppet} - \cite
{vit09x}, sinh-Gordon or Poisson - Boltzmann equation \cite{vit94b},
Lorenz - like systems \cite{vit07}, or water waves \cite{thaker} - \cite{johnson}.
\par
The discussion below will be devoted to the application of the modified method of simplest 
equation for obtaining exact and approximate solutions of nonlinear differential - difference 
equations. The differential-difference equations are much used to describe different 
processes in  complex discrete systems in physics, biology, engineering, etc.. We shall 
discuss below the use of such equations for description of waves in lattices conected to ecological 
food chains.
The method of simplest equation has been established by Kudryashov \cite{kudr05x,kudr05}, 
\cite{kudr08}-\cite{k2} on the basis of  a procedure analogous to the first step
of the test for the Painleve property \cite{hone}. The modified method of simplest 
equation is  simpler for use version of this method \cite{vdk10, vit11, vit09x} where the 
above-mentioned procedure is substituted by  the concept for
the balance equation. Modified method of simplest equation is already  applied
for obtaining exact traveling wave solutions of nonlinear PDEs such as versions of generalized
Kuramoto - Sivashinsky equation, reaction - diffusion  equation, reaction - telegraph 
equation \cite{vdk10}, \cite{vd10}
generalized Swift - Hohenberg equation and generalized Rayleigh equation \cite{vit11},
generalized Fisher equation, generalized Huxley equation \cite{vit09x}, generalized 
Degasperis - Processi equation and b-equation\cite{vit11a}, and to numerous nonlinear PDEs
and ODEs \cite{vit11b}.
\par
The organization of the paper is as follows. In Sect. 2 we define the class of the discussed lattices - the
Riccati lattices. We shall search for Riccati lattices among the members of two classes of lattices: the 
class of generalized Lotka-Volterra lattices and the class of
the generalized Holling lattices. Sect. 3 is devoted to a brief description of the modified method of
simplest equation. In Sect. 4 the method is applied to the differential - difference equations describing
the generalized Lotka - Volterra lattices. It is shown that from this class of lattices only the generalized
Wadati lattice belongs to the class of Riccati lattices. Sect. 5 is devoted to obtaining traveling-wave solutions
of the differential - difference equations that describe lattices from the class of  generalized Holling lattices.
Several concluding remarks are summarized in Sect. 6.
\section{Riccati lattices}
\subsection{Riccati lattices and Riccati equation}
We shall denote as Riccati lattices the class of lattices that admit traveling-wave
solutions obtained by the method of simplest equation on the basis of the use of the
Riccati equation as simplest equation. The equation of Riccati is:
\begin{equation}\label{a1}
\frac{d \Phi}{d \xi} = b^2 - \Phi^2.
\end{equation}
The solution of (\ref{a1})  is
\begin{equation}\label{a2}
\Phi (\xi) = b \tanh [b (\xi + \xi_0)].
\end{equation}
Here, the following notes are in order:
\begin{description}
\item[[1.]]
Another form of the Riccati equation is:
\begin{equation}\label{a3}
\frac{d \tilde{\Phi}}{d \xi} = \tilde{b}^2 - \tilde{a}^2 \tilde{\Phi}^2.
\end{equation}
Eq. (\ref{a3}) has the solution:
\begin{equation}\label{a4}
\tilde{\Phi} (\xi) = \frac{\tilde{b}}{\tilde{a}} \tanh [\tilde{a} \ \tilde{b} \ (\xi + \xi_{0})],
\end{equation}
where $\tilde{a}^2 \tilde{\Phi} (\xi)^{2} < \tilde{b}^{2}$ and $\xi_{0}$ is a constant of integration.
\par
The third form of the Riccati equation is:
\begin{equation}\label{a5}
\frac{d \Psi}{d \xi} = a^* [\Psi (\xi)]^{2} + b^* \Psi (\xi) + c^*,
\end{equation}
which has as a solution
\begin{equation}\label{a6}
\Psi(\xi) = - \frac{b^*}{2 a^*} - \frac{\theta}{2 a^*} \tanh
\left[\frac{\theta (\xi + \xi_0)}{2} \right].
\end{equation}
In Eq. (\ref{a6}) $\theta^{2} = {b^*}^{2} - 4 a^* c^* >0$.
One can easily check that when $\tilde{\Phi}(\xi) = \Psi(\xi) - \frac{b^*}{2 a^*}$ and in addition
$a^* = - \tilde{a}^2$ as well as $\tilde{b}^2 = \frac{4 a^* c^* -{b^*}^2}{4 a^*}$ then the equation (\ref{a5})
is reduced to the Eq. (\ref{a3}) and the solution (\ref{a6}) is reduced to the solution
(\ref{a4}).
\par
The Riccati equation (\ref{a3}) can be further reduced to Eq. (\ref{a1}). Let 
\begin{equation}\label{a7}
\tilde{\Phi} = \cfrac{1}{\tilde{a^2}} \Phi; \hskip.5cm b = \tilde{a} \tilde{b}. 
\end{equation}
The substitution of Eq. (\ref{a7}) in (\ref{a3}) leads to Eq. (\ref{a1}) and the
solution (\ref{a6}) is reduced to the solution (\ref{a2}). We shall use Eq. (\ref{a1}) below
and its solution (\ref{a2}).
\item[[2.]]
The Riccati equation is one of the many possible simplest equations (for other possibilities
see for an example \cite{vit11b}).  Tanh - function and  the Riccati equation are already 
applied to many lattices described by differential - difference equations. Tanh-function was 
used for an example for obtaining exact traveling wave solutions of a nonlinear lattice 
Klein - Gordon model \cite{comt1}. Tanh-function was used also for obtaining of exact  
traveling-wave solution of differential - difference
equation in \cite{bald1} . Exact traveling wave solution are obtained in \cite{bald1}
for the Ablowitz - Ladik lattice, several variants of the non-relativistic and relativistic Toda 
lattice for Volterra lattice, for discretized mKdV lattice and for the hybrid (Wadati) lattice.
Solution of the Riccati equation was used for investigation of the Wadati lattice equation by 
Xie and Wang \cite{xie}. Recently Aslan \cite{a1,a2} used the (G'/G) - method for obtaining 
exact traveling waves of differential - difference equations. Kudryashov \cite{k10a} 
has shown that the 
(G'/G) - method is equavalent to the method of simplest equation for the case when the 
equation of Riccati is used as simplest equation.
\item[[3.]]
The new knowledge this paper adds to the significant amount of research of 
differential-difference equations is as follows: First, we show that from the class of generalized Lotka - 
Volterra lattices discussed below only the Wadati lattice belongs also to the class of 
Riccati lattices. In addition, we discuss exact traveling wave solutions of  the class of 
Holing lattices and show that many of these lattices are Riccati lattices.
\end{description}
\subsection{Generalized Lotka-Volterra lattices. Holling lattices}
Let us consider a chain  of species. The number of each kind of species is 
$M_1,M_2,\dots$. Let us assume that the number of $n$-th species $M_n$ increases by 
collision with $n+1$-th species and decreases by collision with $n-1$-th species. Then
we can write:
\begin{equation}\label{b1}
\frac{d M_n}{d t} = M_n (M_{n+1} - M_{n-1}).
\end{equation} 
Eq. (\ref{b1}) is a simple example of a differential - difference equation that models a
Lotka - Volterra lattice. 
Wadati \cite{wadati} has discussed the class of lattices:
\begin{equation}\label{b2}
\frac{d M_n}{d t} = (\alpha + \beta M_n + \gamma M_n^2)(M_{n+1}-M_{n-1}),
\end{equation}
which generalizes the Lotka - Volterra latttices of kind Eq. (\ref{b1}). 
\par
We shall discuss below the following two generalizations of the Wadati  
and Lotka - Volterra lattices:
\begin{description}
\item[(1.)] Generalized Lotka - Volterra lattices:
\begin{equation}\label{b3}
\frac{d M_n}{d t} = F(M_n)(M_{n+1}-M_{n-1}),
\end{equation}
where $F(M_n)$ is a polynomial of $M_n$ and
\item[(2.)] Generalized Holling lattices:  
\begin{equation}\label{b4}
\frac{d M_n}{d t} = \dfrac{F(M_n)}{G(M_n)}(M_{n+1}-M_{n-1}),
\end{equation}
where $F(M_n)$ and $G(M_n)$ are polynomials of $M_n$.
\end{description}
As we can see Eq. (\ref{b3}) is a straightforward
generalization of the Wadati lattice equation (\ref{b2}).  Eq. (\ref{b4}) reflects the possibility
of Holling functional response in population dynamics \cite{holling}. We shall denote because 
of this the lattices modeled by this equation as generalized Holling lattices. 
\par
We shall study below  the conditions which ensure that the lattices described by
Eqs. (\ref{b3}) and (\ref{b4}) belong to the class of Riccati lattices defined above.
The basis of our investigation will be the modified method of simplest equation
for obtaining exact and approximate solutions of nonlinear PDEs.
\section{The modified method of simplest equation}
Let us have a partial differential equation and let by means of an
appropriate ansatz this equation be reduced to the nonlinear ODE:
\begin{equation}\label{gen_eq}
P \left( F(\xi),\frac{d F}{d \xi},\frac{d^{2} F}{d \xi^{2}},\dots \right) = 0.
\end{equation}
For large class of equations from the kind (\ref{gen_eq}) exact solution 
can be constructed as finite series:
\begin{equation}\label{solution}
F(\xi) = \sum_{\mu=-\nu}^{\nu_1} a_{\mu} [\Phi (\xi)]^{\mu},
\end{equation}
where  $\nu>0$, $\mu >0$, $p_{\mu}$ are  parameters and $\Phi (\xi)$ is a solution of 
some ordinary differential equation
referred to as the simplest equation. The simplest equation is of lesser 
order than (\ref{gen_eq}) and we know the general solution of
the simplest equation or we know at least exact analytical particular
solution(s) of the simplest equation \cite{kudr08, kudr07x}. 
\par
The  modified method of simplest equation can be applied to
nonlinear partial differential equations of the kind:
\begin{equation}\label{basic_eq1}
E \left(\frac{\partial ^{\omega_1} F}{\partial x ^{\omega_1}},
\frac{\partial ^{\omega_2} F}{\partial t ^{\omega_2}}, 
\frac{\partial ^{\omega_3} F}{\partial x ^{\omega_4} \partial t^{\omega_5}}\right) = G(F),
\end{equation}
where $\omega_3 = \omega_4 + \omega_5$ and
\begin{enumerate}
\item
$\frac{\partial ^{\omega_1} F}{\partial x ^{\omega_1}}$ denotes the set of derivatives:
$$
\frac{\partial ^{\omega_1} F}{\partial x ^{\omega_1}} = \left( \frac{\partial F}{\partial 
x}, \frac{\partial^2 F}{\partial x^2}, \frac{\partial F^3}{\partial x^3}, \dots \right).
$$
\item
$\frac{\partial ^{\omega_2} F}{\partial t ^{\omega_2}}$ denotes the set of derivatives:
$$
\frac{\partial ^{\omega_2} F}{\partial t ^{\omega_2}} = \left( \frac{\partial F}{\partial 
t}, \frac{\partial^2 F}{\partial t^2}, \frac{\partial f^3}{\partial t^3}, \dots \right).
$$
\item
$\frac{\partial^{\omega_3} F}{\partial x^{\omega_4} \partial t^{\omega_5}}$ denotes the set
of derivatives:
$$
\frac{\partial^{\omega_3} F}{\partial x^{\omega_4} \partial t^{\omega_5}} = \left( 
\frac{\partial^2 F}{\partial x \partial t}, \frac{\partial^3 F}{\partial x^2 \partial t},
\frac{\partial F^3}{\partial x \partial t^2}, \dots  \right).
$$
\item
$G(F)$ can be:
\begin{enumerate}
\item
polynomial of $F$ or;
\item
function of $F$ which can be reduced to polynomial of $F$ by means of
Taylor series for small values of $F$.
\end{enumerate}
\item
The function $E$ can be an arbitrary sum of products of arbitrary number of its arguments.
Each argument in each product can have arbitrary power. Each of the products
can be multiplied by a function of $F$ which can be:
\begin{enumerate}
\item
polynomial of $F$ or;
\item
function of $F$ which can be reduced to polynomial of $F$ by means of
Taylor series for small values of $F$.
\end{enumerate}
\end{enumerate}
The modified method of simplest equation for this class of equations
allows us in principle to search for:
\begin{enumerate}
\item
Exact traveling-wave solutions of (\ref{basic_eq1}) if $G(F)$ and the multiplication functions
form item 5. above are polynomials; 
\item
Approximate traveling-wave solutions for small $F$ in all other cases.
\end{enumerate}
The application of the modified method of simplest equation is based on the following steps: 
\begin{itemize}
\item
The solved class of NPDE of
kind (\ref{basic_eq1}) is reduced to a class of nonlinear ODEs of the kind (\ref{gen_eq})
by means of an appropriate ansatz (for an example the traveling-wave ansatz); 
\item
The finite-series solution (\ref{solution}) is substituted in (\ref{gen_eq}) and as a result a polynomial
of $\Phi(\xi)$ is obtained. Eq. (\ref{solution}) is a solution of (\ref{gen_eq}) if all coefficients of
the obtained polynomial of $\Phi(\xi)$ are equal to $0$;
\item
One ensures by means of a balance equation that there are at least two terms in the coefficient of the highest
power of $\Phi(\xi)$. The balance equation gives a relationship between the parameters of the solved class of
equations and the parameters of the solution;
\item
The application of the balance equation and the equalizing the coefficients of the polynomial of $\Phi(\xi)$
to $0$ leads to a system of nonlinear relationships among the parameters of the solution and the parameters of
the solved class of equation;
\item
Each solution of the obtained system of nonlinear algebraic equations leads to a solution of  
a nonlinear PDE from the investigated class of  nonlinear PDEs.
\end{itemize}
\section{The uniqueness of the  Wadati lattice}
Let us apply the modified method of simplest equation to the lattice equation (\ref{b3}).
We are interested in traveling waves and introduce the traveling - wave coordinate $\xi_n =
c \ t + d \ n + \xi_0$, where $c$, $d$ and $\xi_0$ are parameters. After the substitution of 
the traveling-wave coordinate in Eq. (\ref{b3}), we obtain the lattice equation:
\begin{equation}\label{e1}
c \frac{d M_n}{d \xi_n} - F(M_n) \  [M_{n+1} - M_{n-1}] =0.
\end{equation}
As we are interested in the Riccati lattices we search the traveling - wave solution of
Eq. (\ref{e1}) as a sum of powers of the solution of the Riccati equation:
\begin{equation}\label{e2}
M_n (\xi_n) = \sum_{k=0}^K a_k [\Phi (\xi_n)]^k; \hskip.5 cm \frac{d \Phi}{d \xi_n} = b^2 -
[\Phi(\xi_n)]^2.
\end{equation}
For the polynomial $F(M_n)$ we assume:
\begin{equation}\label{e3}
F(M_n) = \sum_{l=0}^{L} c_l M_n^l = \sum_{l=0}^L \left[ \sum_{k=0}^K a_k \Phi^k \right]^l.
\end{equation}
The substitution of Eqs. (\ref{e2}), (\ref{e3}) in Eq. (\ref{e1}) leads to the following equation:
\begin{eqnarray}\label{e4}
c [b^2 - \sigma^2 \Phi^2]^K \sum_{k=0}^K (k b^2 a_k \Phi^{k-1} - k a_k \Phi^{k+1})-
\sum_{l=0}^L \left[ \sum_{k=0}^K a_k \Phi^k \right]^l \times \nonumber \\
\Bigg \{ \sum_{k=0}^K a_k b^k [(b + \sigma \Phi)^{K-k} (b - \sigma \Phi)^K (b \sigma + \Phi)^k - (b + \sigma \Phi)^K 
(b- \sigma \phi)^{K-k} (\Phi - b \sigma)^k] \Bigg \} =0,
\nonumber \\
\end{eqnarray}
where $\sigma = \tanh (b \ d)$. 
\par
Let us now derive the balance equation for Eq. (\ref{e4}). The maximum powers
connected to the different groups of terms in Eq. (\ref{e4}) are as follows.
\begin{eqnarray}\label{e5}
{\rm \bf  Term} \ \ c [b^2 - \sigma^2 \Phi^2]^K \sum_{k=0}^K (k b^2 a_k \Phi^{k-1} - k a_k \Phi^{k+1}) &\to& 3 K +1, \nonumber \\
{\rm \bf Term } \ \ \sum_{l=0}^L \left[ \sum_{k=0}^K a_k \Phi^k \right]^l 
 \Bigg \{ \sum_{k=0}^K a_k b^k (b + \sigma \Phi)^{K-k} (b - \sigma \Phi)^K (b \sigma + \Phi)^k \Bigg \} &\to& K L + 2, K \nonumber \\
 {\rm \bf Term} \ \ \sum_{l=0}^L \left[ \sum_{k=0}^K a_k \Phi^k \right]^l 
\Bigg \{ \sum_{k=0}^K a_k b^k  (b + \sigma \Phi)^K (b- \sigma \phi)^{K-k} (\Phi - b \sigma)^k \Bigg \} 
&\to& K L + 2 K. \nonumber \\
\end{eqnarray}
Thus we have two possibilities for balance equation:
\begin{itemize}
\item
Balance between the first and the second term in Eq. (\ref{e5}):
\begin{equation}\label{e6}
K + 1 = K L.
\end{equation}
\item
Balance between the second and the third term in Eq. (\ref{e5}):
\begin{equation}\label{e7}
K L + 2 K = K L + 2 K.
\end{equation}
\end{itemize}
As $K$ and $L$ must be integers and from Eq. (\ref{e6}) $L=1 + \dfrac{1}{K}$ the balance (\ref{e6}) is valid only for 
the case $K=1$, $L=2$. In all other cases the balance has to be (\ref{e7}). 
\par
It is easily to see that the balance equation (\ref{e7}) is not acceptable. The application of this balance equation 
to Eq. (\ref{e4}) leads to terms of the kind:
$$
\sigma^{2K -k} \Phi^{2K} [(-1)^K - (-1)^{K-k}], \hskip.5cm k=0,1,\dots, K.
$$
Except for the case $K=0$ in all other cases terms arise for which 
$[(-1)^K - (-1)^{K-k}] \ne 0$. This  fact requires $\sigma =0$ which leads to
$d=0$ which is not acceptable for the discussed problem. Below we
shall discuss because of this the balance equation (\ref{e6}).
\par
As we have mentioned above Eq. (\ref{e6}) leads to $K=1$ and $L=2$ which is exactly
the case of the generalized Wadati lattice.  The application of the modified method of
simplest equation to Eq. (\ref{e4}) with $K=1$ and $L=2$ leads to the following
system of 5 nonlinear algebraic relationships among the parameters 
of the equation and the parameters of the solution:
\begin{eqnarray}\label{e8}
\sigma a_1 [c \sigma + 2 c_2 a_1^2 b] &=& 0, \nonumber \\
2 \sigma  a_1^2 b [c_1  + 2 c_2 a_0 ] &=& 0, \nonumber \\
a_1 b [-b c (1+\sigma^2) + 2 \sigma (c_0 + c_1 a_0 + c_2 a_0^2 - 2 c_2 a_1^2 b^2)] &=& 0, \nonumber \\
2 \sigma a_1^2 b^3  [c_1 + 2 c_2  a_0]   &=& 0, \nonumber \\
a_1 b^3 [b c   - 2 \sigma (c_0 + c_1 a_0 + c_2  a_0^2) ] &=& 0. 
\end{eqnarray}
One solution of this system is:
\begin{equation}\label{e9}
c = \frac{\sigma (4 c_0 c_2 - c_1^2)}{2 b c_2}; \hskip.25cm
a_0 = - \frac{c_1}{2 c_2}; \hskip.25 cm a_1 = \frac{\sigma \sqrt{c_1^2 - 4 c_0 c_2}}{2 b c_2},
\end{equation}
and the corresponding solution of Eq. (\ref{e4}) is
\begin{eqnarray}\label{e10}
M_n (\xi_n) =  - \frac{c_1}{2 c_2}  + \frac{\tanh (b d) \sqrt{c_1^2-4 c_0 c_2 }}{2 c_2} \tanh 
\left[  b \left( - \frac{(c_1^2 - 4 c_0 c_2) \tanh (b \ d) \ t}{2 b c_2} + d\ n + \xi_0 \right)  \right].
\nonumber \\
\end{eqnarray}
The  solution (\ref{e10}) has been  obtained by different authors. The interesting point  is that the 
discussion of the possible balance equations above has shown the unique position of the Wadati lattice as the 
only Riccati lattice of the kind (\ref{e2}) from the class of the
generalized Lotka - Volterra lattices discussed here.
\section{Holling lattices}
To the best of our knowledge the class of generalized Holling lattice equations was not discussed up to
now. Thus the obtained below traveling-wave solutions are new.
\par
Let us now discuss Eq. (\ref{b4}) where $F(M_n$) be the same as in (\ref{e3}) ,$M_n$ be
given by Eq. (\ref{e2}). In addition let  $G(M_n)$ be:
\begin{equation}\label{o1}
G(M_n) = \sum_{p=0}^P d_p M_n^p = \sum_{p=0}^P d_p \left[\sum_{k=0}^K a_k \Phi^k \right]^p.
\end{equation}
The substitution of Eqs. (\ref{e2}), (\ref{e3}) and (\ref{o1}) in Eq. (\ref{b4})
and the switching to the traveling-wave coordinate leads to the equation:
\begin{eqnarray}\label{o2}
c [b^2 - \sigma^2 \Phi^2]^K \Bigg[ \sum_{p=0}^P d_p \left( \sum_{k=0}^K  a_k \Phi^k \right )^p \Bigg ] \sum_{k=0}^K (k b^2 a_k \Phi^{k-1} - k a_k \Phi^{k+1})-
\sum_{l=0}^L \left[ \sum_{k=0}^K a_k \Phi^k \right]^l \times \nonumber \\
\Bigg \{ \sum_{k=0}^K a_k b^k [(b + \sigma \Phi)^{K-k} (b - \sigma \Phi)^K (b \sigma + \Phi)^k - (b + \sigma \Phi)^K (b- \sigma \phi)^{K-k} (\Phi - b \sigma)^k] \Bigg \} =0.
\nonumber \\
\end{eqnarray}
Here, we again have two possibilities for balance equations:
\begin{equation}\label{o3}
K  P + K + 1 = K  L,
\end{equation}
and
\begin{equation}\label{o4}
K  L + 2 K = K  L + 2 K.
\end{equation}
The balance equation (\ref{o4}) as in previous section leads to $d=0$ which is unacceptable
for the discussed problem. Thus we shall work on the basis of the
balance equation (\ref{o3}). We note that when $P=0$ Eq. (\ref{o3}) reduces to the balance 
equation (\ref{e6}) from the previous section. In addition from Eq. (\ref{o3}) we obtain
$L = P + 1 + \dfrac{1}{K}$. As $L$, $P$ and $K$ must be integer then we must set $K=1$
in Eq. (\ref{o3}). We shall discuss below the simplest cases $P=1$, $P=2$ and $P=3$. 
\subsection{Case $P=1$, $L=3$}
For this case Eq. (\ref{b4}) becomes: 
\begin{equation}\label{p0}
\frac{d M_n}{d t} = \dfrac{c_0+ c_1 M_n + c_2 M_n^2 + c_3 M_n^3}{d_0 + d_1 M_n}(M_{n+1}-M_{n-1}).
\end{equation}
The application of the modified method of simplest equation reduces Eq. (\ref{p0}) to
the following system of nonlinear algebraic relationships:
\begin{eqnarray}\label{p1}
\sigma a_1^2 [ c \sigma d_1 + 2 c_3 a_1^2 b] &=& 0, \nonumber \\
\sigma a_1 [c \sigma (d_0 + d_1 a_0)  +  2 a_1^2  b (c_2  + 3 c_3  a_0 )] &=& 0, \nonumber \\
a_1^2 b [2 \sigma (2 c_2 a_0 + c_1 + 3 c_3 a_0^2) - b( 2 c_3 a_1^2 b \sigma - c d_1 - c \sigma^2 d_1)] &=& 0, 
\nonumber \\
a_1 b \{\sigma [ 2 (c_0 + c_1 a_0 + c_3 a_0 ^3 + c_2 a_0^2) - 2 b^2 (c_2 a_1^2 + 3 c_3 a_0 a_1^2)] - b c [ (d_0 + d_1 a_0)- && \nonumber \\
\sigma^2 (d_0 + d_1 a_0)] \} &=& 0, \nonumber \\
a_1^2 b^3 [b c  d_1 - 2 \sigma  (2 c_2 a_0  + c_1 + 3 c_3 a_0^2)] &=& 0, \nonumber \\
 a_1 b^3[ b c (d_0 + d_1 a_0) - 2 \sigma  (c_0 + c_1 a_0 + c_3 a_0^3 + c_2 a_0^2)] &=& 0. 
\nonumber \\
\end{eqnarray}
One solution of this system is:
\begin{eqnarray}\label{p2}
c &=& -\frac{\sigma (2 c_2 c_3 d_0 d_1 + c_2^2 d_1^2 - 
4 c_1 c_3 d_1^2 - 3 c_3^2 d_0^2)}{2 b c_3 d_1^3},  \nonumber \\
a_1 &=& \frac{\sigma \sqrt{2 c_2 c_3 d_0 d_1 + c_2^2 d_1^2 - 
4 c_1 c_3 d_1^2 - 3 c_3^2 d_0^2}}{2 b c_3 d_1}, \nonumber \\
a_0 & = & - \frac{c_2 d_1 - c_3 d_0}{2 c_3 d_1}; \hskip.5cm
c_0 = \frac{d_0 (c_1 d_1^2 + c_3 d_0^2 - c_2 d_0 d_1)}{d_1^3}.
\end{eqnarray}
and the corresponding traveling-wave is:
\begin{eqnarray}\label{p3}
M_n (\xi_n) =  - \frac{1}{2 c_3 d_1} \Bigg \{ c_2 d_1 - c_3 d_0 +\frac{\tanh(b \ d)}{d} \sqrt{2 c_2 
c_3 d_0 d_1 + c_2^2 d_1^2 - 4 c_1 c_3 d_1^2 - 3 c_3^2 d_0^2} \times \nonumber \\
\tanh \Bigg[  -\frac{\tanh (b \ d) (2 c_2 c_3 d_0 d_1 + c_2^2 d_1^2 - 
4 c_1 c_3 d_1^2 - 3 c_3^2 d_0^2)}{2  c_3 d_1^3} t  + d \ n + \xi_0 \Bigg] \Bigg \}. \nonumber \\
\end{eqnarray}
\subsection{Case $P=2$, $L=4$}
For this case Eq. (\ref{b4}) becomes: 
\begin{equation}\label{q0}
\frac{d M_n}{d t} = \dfrac{c_0+ c_1 M_n + c_2 M_n^2 + c_3 M_n^3 + c_4 M_n^4}{d_0 + d_1 M_n + d_2 M_n^2}
(M_{n+1}-M_{n-1}).
\end{equation}
The application of the modified method of simplest equation reduces Eq. (\ref{q0}) to
a system of 7 nonlinear algebraic
relationships among the parameters of the equation and the parameters of the solution.
One solution of this nonlinear algebraic system is:
\begin{eqnarray}\label{q1}
a_0 &=& \frac{c_4 d_1 - c_3 d_2}{2 c_4 d_2}, \nonumber \\
a_1 &=&  \frac{\sigma}{2 b c_4 d_2} \sqrt{\frac{4 c_2 d_2^2 c_4^2 d_1 - 4 c_2 d_2^3 c_4 c_3 - 3 d_2 c_3 c_4^2 d_1^2 + 
d_2^3 c_3^3 + 4 c_1 c_4^2 d_2^3 + 2 c_4^3 d_1^3}{c_3 d_2 - 2 c_4 d_1}},
\nonumber \\
c &=& - \frac{\sigma}{2 b c_4 d_2^3 (c_3 d_2 - 2 c_4 d_1) } [4 c_2 d_2^2 c_4^2 d_1 - 4 c_2 d_2^3 c_4 c_3 - 3 d_2 c_3 
c_4^2 d_1^2 + d_2^3 c_3^3 + 4 c_1 c_4^2 d_2^3 + 2 c_4^3 d_1^3],
\nonumber \\
c_0 &=& - \frac{1}{d_2^4 (2 c_4 d_1 - c_3 d_2 )^2} \Bigg[d_1^3 d_2^3 c_3^3 + c_3^2 c_1 d_1 d_2^5 - 2 c_3^2 c_2 d_1^2 
d_2^4 - 3 c_3^2 c_4 d_1^4 d_2^2 - c_3 c_1 c_2 d_2^6 + 
\nonumber \\
&& c_3 d_2^5 c_2^2 d_1 - c_3 c_4 d_1^2 c_1 d_2^4 + 4 c_3 c_2 c_4 d_1^3 d_2^3 + 3 c_3 c_4^2 d_1^5 d_2 + c_1^2 c_4 d_2^6 
- d_2^4 c_2^2 c_4 d_1^2 - \nonumber \\ 
&& 2 c_2 c_4^2 d_1^4 d_2^2-c_4^3 d_1^6 \Bigg], \nonumber \\
d_0 &=& \frac{c_3 d_1^2 d_2 - c_2 d_1 d_2^2 + c_1 d_2^3 -c_4 d_1^3}{d_2 (c_3 d_2 - 2 c_4 d_1)}.
\nonumber \\
\end{eqnarray}
The corresponding traveling-wave is:
\begin{eqnarray}\label{q2}
M_n (\xi_n) &= & \frac{1}{2 c_4 d_2} \Bigg \{ c_4 d_1 - c_3 d_2 +\nonumber \\
&& \tanh( b \ d) \sqrt{\frac{4 c_2 d_2^2 c_4^2 d_1 - 4 c_2 d_2^3 c_4 c_3 - 3 d_2 c_3 c_4^2 d_1^2 + d_2^3 c_3^3 + 4 c_1 
c_4^2 d_2^3 + 2 c_4^3 d_1^3}{c_3 d_2 - 2 c_4 d_1}} \times \nonumber \\
&& \tanh \Bigg[  - \frac{b \tanh(b \ d)}{2 b c_4 d_2^3 (c_3 d_2 - 2 c_4 d_1) } (4 c_2 d_2^2 c_4^2 d_1 - 4 c_2 d_2^3 
c_4 c_3 - 3 d_2 c_3 c_4^2 d_1^2 + d_2^3 c_3^3 + \nonumber \\
&& 4 c_1 c_4^2 d_2^3 + 2 c_4^3 d_1^3)\ t +
d \ n + \xi_0 \Bigg ] \Bigg \}. \nonumber \\
\end{eqnarray}
\subsection{Case $P=3$, $L=5$ }
For this case Eq. (\ref{b4}) becomes: 
\begin{equation}\label{r0}
\frac{d M_n}{d t} = \dfrac{c_0+ c_1 M_n + c_2 M_n^2 + c_3 M_n^3 + c_4 M_n^4 + c_5 M_n^5}{d_0 + d_1 M_n + d_2 M_n^2 + 
d_3 M_n^3}(M_{n+1}-M_{n-1}).
\end{equation}
The application of the modified method of simplest equation reduces Eq. (\ref{r0}) to
a system of 8 nonlinear algebraic
relationships among the parameters of the equation and the parameters of the solution.
One solution of this nonlinear algebraic system is:
\begin{eqnarray}\label{r1}
c &=& \frac{\sigma (3 c_5^2 d_2^2 - 2 c_5 d_2 d_3 c_4 - 4 d_1 d_3  c_5^2 - d_3^2 c_4^2 + 4 c_3 c_5 d_3^2) }{2 b c_5 
d_3^3}, \nonumber \\
a_0 &=&  \frac{c_5 d_2 -c_4 d_3}{2 c_5 d_3}, \nonumber \\
a_1 &=& \frac{\sigma \sqrt{-3 c_5^2 d_2^2 + 2 c_5 d_2 d_3 c_4 + 4 d_1 d_3 c_5^2 + d_3^2 c_4^2 - 4 c_3 c_5 d_3^2}}{2 b 
c_5 d_3}, \nonumber \\
d_0 &=& \frac{c_0 d_3^3}{c_3 d_3^2 - d_3 c_4 d_2 - d_3 d_1 c_5 + d_2^2 c_5}, \nonumber \\
c_1 &=& \frac{C_1}{d_3^3 (c_3 d_3^2 - d_3 c_4 d_2 - d_3 d_1 c_5 + d_2^2 c_5)}, \nonumber \\
C_1 &=& -2 d_1 d_3^3 c_3 c_4 d_2 - 2 c_5^2 d_2^2 d_1^2 d_3 + c_5^2 d_2^4 d_1 - 2 c_5 d_2^3 d_3 c_4  d_1 + 2 c_5 d_2^2 
d_1 d_3^2 c_3 + d_2^2 d_3^2 c_4^2 d_1 + \nonumber \\
&& 2 c_5 d_2 d_1^2 d_3^2 c_4 + d_1^3 d_3^2 c_5^2 + d_1 d_3^4 c_3^2 - 2 d_1^2 d_3^3 c_5 c_3 - c_5 d_2 c_0 d_3^4 + d_3^5 
c_4 c_0, \nonumber \\
c_2 &=& \frac{C_2}{d_3^3 (c_3 d_3^2 - d_3 c_4 d_2 - d_3 d_1 c_5 + d_2^2 c_5)}, \nonumber \\
C_2 &=& c_5^2 d_2^5 + c_0 c_5 d_3^5 + 4 c_5 d_2^2 d_1 d_3^2 c_4 - d_2 d_3^3 c_4^2 d_1 - 2 c_3 d_3^3 d_2^2 c_4 - 3 d_1 
d_3^3 c_5 c_3 d_2 + \nonumber \\
&& d_1 d _3^4 c_3 c_4 - 2 c_5 d_2^4 d_3 c_4 + d_2^3 d_3^2 c_4^2 - 3 c_5^2 d_2^3 d_1 d_3 + 2 c_3 c_5 d_3^2 d_2^3 + 2 
d_1^2 d_3^2 c_5^2 d_2 - d_1^2 d_3^3 c_5 c_4 c_3^2 d_3^4 d_2. \nonumber \\
\end{eqnarray}
The corresponding traveling wave is:
\begin{eqnarray}\label{r2}
M_n (\xi_n) &=& \frac{1}{2 c_5 d_3} \Bigg \{ c_5 d_2 -c_4 d_3 + \tanh(b \ d) \sqrt{-3 c_5^2 d_2^2 + 2 c_5 d_2 d_3 c_4 
+ 4 d_1 d_3 c_5^2 + d_3^2 c_4^2 - 4 c_3 c_5 d_3^2} \times \nonumber \\
&& \tanh \Bigg[ \Bigg( \frac{\tanh(b \ d) (3 c_5^2 d_2^2 - 2 c_5 d_2 d_3 c_4 - 4 d_1 d_3  c_5^2 - d_3^2 c_4^2 + 4 c_3 
c_5 d_3^2) t }{2  c_5 d_3^3}  + d \ n + \xi_0\Bigg ) \Bigg] \Bigg \}. \nonumber \\
\end{eqnarray}
The same procedure can be continued for $P=4,5,\dots$ and 
the differential-difference equations for the 
corresponding Holling lattices will be reduced to a nonlinear algebraic systems consisting of 
$9,10,\dots$ equations  as a result of the application of the 
modified method of simplest equation. Each solution  
will lead to a traveling wave constructed on the basis of Riccati equation
if we are able  to solve these nonlinear systems. 
\section{Concluding remarks}
Lattices have many applications in mathematics and physics. This is one of the reason for the
importance of the differential - difference equations that often are used to model wave 
processes in lattices connected to physical chemical or biological systems.  In this paper we 
have applied the modified method of simplest equation for identification of the Riccati 
lattices among the classes of the generalized Lotka - Volterra lattices  and generalizing 
Holling lattices. The analysis of the balance equation arising from the application of the 
method of simplest equation has shown that the Wadati lattice is unique in the
class of the generalized Lotka - Volterra lattice as it is the only Riccati lattice of class (
\ref{e2}) among the lattices of the generalized Lotka - Volterra class. Many more Riccati 
lattices can be found in the class of generalized Holling lattices. We have obtained exact 
traveling wave solutions for the simplest three Riccati lattices that are Holling lattices too.
\par
The identification of the Riccati lattices is important task as the connected to these
lattices waves of $\tanh$-kind describe a kind of switching between the states in the
corresponding lattice. The presence of Riccati lattices among the lattice models used
in different scientific areas shows that probably this kind of switching between the states
is a frequently arising phenomenon and fundamental property of a large class of natural
systems. This paper is a first step from a future research on identifying and studying
the properties of the Riccati lattices.   

\end{document}